\documentclass[mathleft
]{an}
\usepackage{graphicx}
\usepackage{times}
\begin{document}

% The following seven commands are intended for editorial usage and should be ignored by
% the author(s).
\Pagespan{789}{}% Document's page range. 
% If second parameter is left empty, the last page is computed automatically.
\Yearpublication{2006}%
\Yearsubmission{2005}%
\Month{11}%   
\Volume{999}%  
\Issue{88}% 
% \DOI{This.is/not.aDOI}% 

\title{A Short Review of Relativistic Iron Lines from Stellar-Mass Black Holes}

\author{J. M. Miller\inst{1}}
%Example 
%for footnote, note the usage of the \texttt{fnmsep}
%command as separator between institute number and footnote mark} 
%\and  G.H. Ostwriter\inst{2,3
\titlerunning{Fe K Lines in Black Holes}
\authorrunning{J. M. Miller}
\institute{University of Michigan Department of Astrophysics, 500
Church Street, Ann Arbor MI 48109-1042 USA, jonmm@umich.edu}

\received{01 Sept 2006}
\accepted{11 Oct 2006}
\publonline{later}

\keywords{X-rays: binaries -- Accretion, accretion disks}

\abstract{%
In this contribution, I briefly review recent progress in detecting
and measuring the properties of relativistic iron lines observed in
stellar-mass black hole systems, and the aspects of these lines that
are most relevant to studies of similar lines in Seyfert-1 AGN.  In
particular, the lines observed in stellar-mass black holes are not
complicated by complex low-energy absorption or partial-covering of
the central engine, and strong lines are largely independent of the
model used to fit the underlying broad-band continuum flux.  Indeed,
relativistic iron lines are the most robust diagnostic of black hole
spin that is presently available to observers, with specific
advantages over the systematics--plagued disk continuum.  If accretion
onto stellar-mass black holes simply scales with mass, then the
widespread nature of lines in stellar-mass black holes may indicate
that lines should be common in Seyfert-1 AGN, though perhaps harder to
detect.}

\maketitle

\section{Introduction}

Broad Fe~K emission lines plausibly arising in the inner disk around
black holes, and shaped by the strong Doppler shifts and gravitational
redshifts expected there, were first reported in stellar-mass black
holes (Barr, White, \& Page 1985; van der Woerd, White, \& Kahn 1989).
Though these detections were achieved with low--resolution gas
spectrometers, their discovery was met with enthusiasm, and
theoretical papers showing their reported widths to be consistent with
extreme dynamical broadening soon followed (Fabian et al.\ 1989, Laor
1991).  It was also quickly realized that Fe~K emission lines are
merely the most prominent feature of the overall response of an
accretion disk to external irradiation by hard X-ray flux (George \&
Fabian 1991).

Although strong Fe~K lines in Seyfert AGN were well-known at the time
broad lines were found in stellar-mass black holes, the lower flux in
AGN lines made it difficult to measure their shape.  In the {\it
ASCA} era, however, studies of broad iron lines in Seyfert-1 AGN advanced
rapidly.  The SIS (solid-state imaging spectrometer, an X-ray CCD
spectrometer) aboard {\it ASCA} revealed that the lines in some
Seyfert-1 AGN were not only broad, but asymmetric --- consistent with
strong Doppler shifts and gravitational redshifts expected in the
inner accretion disk (see, e.g. Tanaka et al.\ 1995; Iwasawa et al.\
1996; Nandra et al.\ 1997a, Nandra et al.\ 1996b).  Although the SIS
was well-suited to studies of the line profile in Seyfert-1 AGN,
ruinous photon pile-up resulted when stellar-mass black holes were
observed, greatly complicating efforts to recover robust spectra for
detailed line measurements (see, e.g., Ebisawa et al.\ 1996).

In the {\it XMM-Newton} and {\it Chandra} era, considerable effort has
again been devoted to the study of relativistic iron lines.  While
relativistic lines appear to be confirmed in some cases, complex
low-energy absorption or alternative continuum spectral models
initially cast some doubt on relativistic lines in other cases (see,
however, the contributions by Fabian et al.\, Nandra et al.\, and
Reeves et al.\ in these proceedings).  In the stellar-mass black hole
regime, the fast read-out modes and advanced spectrometers aboard {\it
XMM-Newton} and {\it Chandra} have made it possible to measure the
properties of broad iron lines without photon pile-up.  A number of
asymmetric lines have been revealed, showing that they are also shaped
by relativistic blurring at the inner accretion disk.  In this sense,
{\it XMM-Newton} and {\it Chandra} have done for stellar-mass black
holes what {\it ASCA} did for Seyfert-1 AGN.

In Section 2, recent observations of relativistic Fe~K lines in
stellar-mass black holes are reviewed.  In Section 3, the robustness
of relativistic Fe~K lines and their utility as spin diagnostics is
critically examined.  In Section 4, different methods of constraining
black hole spin are discussed.  Finally, the relevance of relativistic
lines in stellar-mass black holes to studies of Seyfert-1 AGN and
objectives for the future studies of relativistic lines are discussed
in Section 5.

\section{Recent Observations of Relativistic Lines}

The first clear detection of a relativistic Fe~K emission line at
moderate or high resolution was made in a {\it Chandra} High Energy
Transmission Grating Spectrometer (HETGS) observation of Cygnus X-1
(Miller et al.\ 2002a).  The resolution of the HETGS permitted a
narrow iron line, either due to illumination of the companion wind or
outer disk, to be separated from an underlying broad, red-shifted
emission line.  The observed line profile does not require a high
degree of black hole spin.  Cygnus X-1 was in an ``intermediate'' flux
state when it was observed with {\it Chandra}; at higher flux levels,
stronger line emission is expected, and more sensitive spectra can be
obtained.  More recent and more sensitive {\it XMM-Newton}/EPIC-pn
observations of Cygnus X-1 confirm the broad line profile at CCD
resolution (see Figure 1; Miller et al.\ 2006, in prep).

\begin{figure}
\includegraphics[width=55mm,height=83mm,angle=-90]{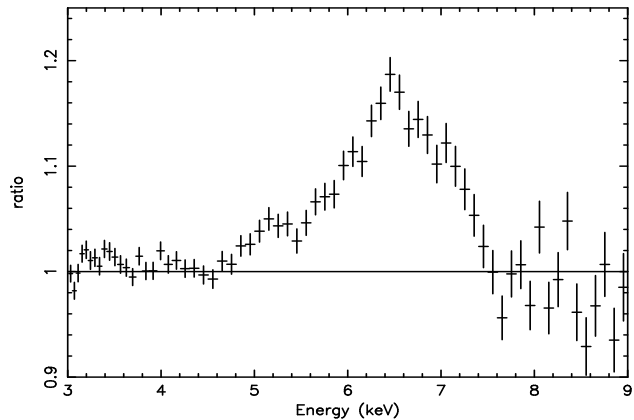}
\caption{The data/model ratio above shows the relativistic Fe~K
emission line profile revealed after fitting a phenomenological model
to a recent {\it XMM-Newton} observation of Cygnus X-1 in a high flux
state (Miller et al.\ 2006, in prep.).  Fits with relativistic line
models indicate that Cygnus X-1 may harbor a black hole with low or
moderate spin, rather than high spin like many black holes with much
older companion stars.  This suggests that accretion may be more
important than black hole formation events in driving spin
parameters.}
\label{label1}
\end{figure}

Soon after the {\it Chandra}/HETGS observation of Cygnus X-1, a new
black hole transient -- XTE J1650$-$500 -- was observed with the {\it
XMM-Newton}/EPIC-pn camera.  This observation revealed a strongly
skewed Fe~K emission line profile, and fits with the Laor line model
imply that the disk extends to within $2~GM/c^{2}$, requiring a black
hole spin parameter in excess of $a \simeq 0.9$ (Miller et al.\
2002b).  Importantly, the line profile and overall disk reflection
spectrum is well fitted with a model that is remarkably similar to
that which describes the same components in an {\it XMM-Newton}
observation of the Seyfert-1 galaxy MCG--6-30-15 (Fabian et al.\
2001).

The relativistic line in XTE J1650$-$500 was confirmed in observations
made by {\it BeppoSAX}, and the observed line variability is
consistent with the predictions of a variability model based on
gravitational light bending near to a spinning black hole (Miniutti,
Fabian, \& Miller 2004).  The same pattern of variability was later
reported in an analysis of {\it RXTE} spectra (Rossi et al.\ 2005).
It is again important to note that this pattern of variability has
also been observed in the Seyfert-1 galaxies (MCG--6-30-15: Miniutti
et al.\ 2003; 1H 0419$-$577: Fabian et al.\ 2005; NGC 4051: Ponti et
al.\ 2006).

An {\it XMM-Newton} observation of the Galactic black hole candidate
SAX~J1711.6$-$3808 revealed another broad Fe~K emission line (in 't
Zand et al.\ 2002a).  The analysis of the {\it XMM-Newton} data was
complicated by photon pile-up, but the line is confirmed in
simultaneous {\it BeppoSAX} spectra.  The line profile is more
symmetric than many others, but is consistent with being shaped by
Doppler shifts and gravitational red-shifts at the inner accretion
disk.

GX 339$-$4 has been a particularly important source for recent studies
of relativistic iron lines in stellar-mass black holes.  As with
Cygnus X-1, a {\it Chandra}/HETGS observation made in an intermediate
flux state first clearly revealed a relativistic line in this source
(Miller et al.\ 2004a).  Fits to the {\it Chandra} spectrum with a
relativistic line and disk reflection model suggest that the disk
extended to within $3~GM/c^{2}$, again requiring a high black hole
spin parameter (Miller et al.\ 2004b).  A long observation made with
the {\it XMM-Newton}/EPIc-pn camera in a higher flux state discovered
the most skewed Fe~K emission line yet found in a stellar-mass black
hole: from a centroid energy consistent with Fe XXV (6.70 keV) and/or
Fe~XXVI (6.97 keV), the red wing extends down to approximately 3 keV,
requiring $r \leq 2~GM/c^{2}$ or $a \geq 0.9$.  Again, the line
profile and the overall continuum spectrum are remarkably similar to
the most extreme line profiles detected in MCG--6-30-15 (e.g. Fabian
et al.\ 2001), and the spectrum can be fit with very similar models.
Most recently, a very long observation of GX~339$-$4 in a low flux
state revealed a relativistic line at a low mass accretion rate,
suggesting that the disk remains close to the black hole even at low
accretion rates (Miller et al.\ 2006a).

Strong Fe~K emission lines are not frequently observed in soft
spectral states, consistent with the expectation that a strong
external source of hard X-rays is needed to irradiate the disk.  Even
in some hard phases, however, lines can appear to be weak, and/or too
narrow to be produced in a disk that extends very close to the black
hole.  An {\it XMM-Newton} observation of the
``microquasar'' GRS~1915$+$105 revealed a broad line that is
inconsistent with an origin in a disk close to the black hole in this
system (Martocchia et al.\ 2006).  This line shape is inconsistent
with profiles observed previously with {\it BeppoSAX} when the source
was in different states.  However, in other phases, the line profile in
GRS~1915$+$105 is consistent with originating close to the black hole
(Martocchia et al.\ 2002, Miller et al.\ 2004c).

Most recently, a skewed Fe~K line was detected in long {\it
XMM-Newton} observations of GRO~J1655$-$40 (Diaz Trigo et al.\
2006).  The source was observed in a high flux state, similar to the
phase in which an extremely skewed line was found in GX~339$-$4.  GRO
J1655$-$40 is an extraordinary system in which independent evidence of
black hole spin has been reported, e.g. from high frequency
quasi- periodic oscillations (QPOs; see Strohmayer 2001).  The
line profile revealed with {\it XMM-Newton} also signals a black hole
with a high spin parameter: fits with a relativistic line model again
signal an inner disk radius within $2~GM/c^{2}$, corresponding to $a
\geq 0.9$.

In general, Fe~K emission lines have been revealed in deep exposures
of black holes in sufficiently bright and hard spectral states, made
with a moderate (CCD) or high (gratings) resolution spectrometers.
However, exceptions exist among highly obscured and high-inclination
systems.  Observations of the black hole
candidates 4U~1630$-$472 and H~1743$-$322 have not revealed broad Fe~K
emission lines, though blue-shifted Fe absorption lines have been
discovered (Miller et al.\ 2006b).

The success of {\it XMM-Newton} and {\it Chandra} observations of
relativistic Fe~K lines in stellar mass black holes bolsters
detections claimed in spectra obtained with low resolution gas
spectrometers.  It has also motivated both new observations and
archival analysis with gas spectrometer data.  Apart from the line in
GRS~1915$+$105 and SAX~J1711.6--3808 discussed above (Martocchia et
al.\ 2002; in 't Zand et al.\ 2002), {\it BeppoSAX} also observed
broad Fe~K lines in Cygnus X-1 (Frontera), XTE J1908+094 (in 't Zand
2002b), and V4641~Sgr (Miller et al.\ 2002c).  An analysis of archival
{\it ASCA} gas imaging spectrometer (GIS) data revealed detections of
broad, relativistic lines in XTE~J1550--564, GRO J1655--40,
GRS~1915$+$105, and Cygnus X-1 (Miller et al.\ 2004c).  Observations
with {\it RXTE}, though it is not a focusing telescope, have detected
broad Fe~K lines in all of the systems mentioned above, and many
others.  {\it RXTE} has been particularly important to the detection
of broad Fe~K lines in black holes, as it is extremely well suited to
defining the broad-band X-ray continuum in the 3--200~keV band.

\section{On the Robustness of Fe~K Lines}

Studies of relativistic lines in Seyfert-1 galaxies have some specific
advantages over comparable studies in stellar-mass black holes.  The
massive black holes in Seyfert galaxies allow one to study dynamical
timescales that are inaccessible in stellar-mass black holes,
providing an improved view of the innermost region near to the black
hole.  Galactic stellar-mass black holes, however, are brighter by
virtue of their proximity, yielding much higher signal-to-noise.
Moreover, some intrinsic differences make it possible to verify that
relativistic Fe~K lines in Seyfert-1 galaxies are likely to be robust.

Low-energy curvature due to warm absorption complicates studies of
relativistic emission lines in some Seyfert galaxies (e.g., NGC 3783).
At least in the case of MCG--6-30-15, rigorous work shows that the
broad line is independent of low-energy curvature (e.g., Vaughan \&
Fabian 2004; Youg et al.\ 2005).  Low energy absorption is unimportant
in stellar-mass black holes.  For instance, the column density in O
VII and O VIII disk wind absorption lines detected in GX~339$-$4
simultaneously with a relativistic Fe~K emission line (Miller et al.\
2004a) is 100 times lower than the column density detected in the same
absorption lines in MCG--6-30-15 (Lee et al.\ 2001), which is known to
have moderate low-energy absorption.  More importantly, the low energy
X-ray emission in stellar-mass black holes is typically dominated by
thermal emission from the disk, which causes a spectral curvature that
is the opposite of what absorption would cause.

It is sometimes argued that the multiple continuum flux components
required to describe stellar-mass black holes might falsely
create the need for a broad line.  This concern is particularly
important when additive components (e.g., from a phenomenological disk
plus power-law model) cross near to the Fe~K range.  An archival {\it
ASCA} spectrum of the black hole candidate 4U~1630$-$472 is an
excellent example that additive components do not easily create the
need for an emission line (see Figure 2).  When the spectrum is fitted
with additive disk and power-law components that cross at
about 6~keV, but there is no evidence for a broad emission line.

\begin{figure}
\includegraphics[width=55mm,height=83mm,angle=-90]{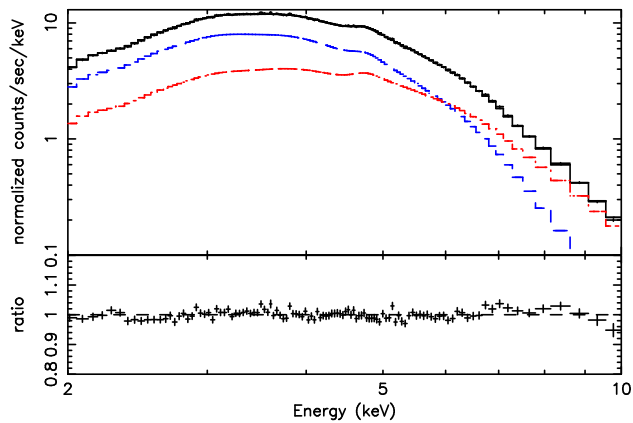}
\caption{An archival {\it ASCA} spectrum of the black hole candidate
4U 1630$-$472 is shown above.  The spectrum was fit with additive
components, including a disk model (shown in blue) and a power-law
model for the hard flux.  The components clearly cross at $\sim6$~keV,
but there is no indication for a broad Fe~K emission line in the
data/model ratio.  The need for two or more continuum flux components
does not act to create false broad or relativistic Fe~K emission lines
in stellar-mass black hole spectra.}
\label{label1}
\end{figure}

Given that relativistic lines can span 2--3~keV in width at their
base, accurate fits to the line (and, therefore, accurate spin
constraints) depend on defining the underlying continuum spectrum
well.  It is possible to confuse this need with the idea that there is
only one correct continuum model, and to worry that relativistic Fe~K
lines cannot be robust if different continuum models describe the
continuum equally well.  In spectra with strong Fe~K lines and a
well-defined (not perfectly defined) continuum, the line profile is
largely independent of the continuum model assumed.  Four fits to the
continuum spectrum of GX~339$-$4 are shown in Figure 3 (Miller et al.\
2006a), and the line profile is clearly robust against these
differences.  In spectra with a low signal-to-noise ratio, the choice
of continuum model can affect the parameters one obtains when fitting
a relativistic line profile, but this is not the case in good spectra.

\begin{figure}
\includegraphics[width=83mm,height=68mm]{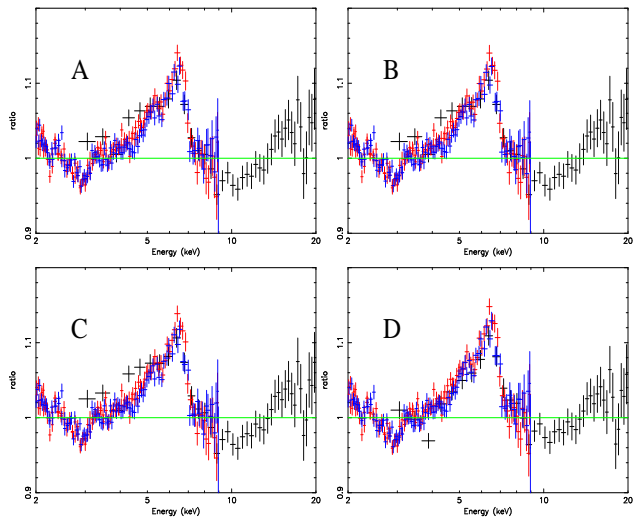}
\caption{The relativistic line profile revealed in simultaneous {\it
XMM-Newton} (red and blue spectra) and {\it RXTE} (black spectra)
observations of GX~339$-$4 is one among several that are robust
against the particular choice of continuum model.  Panel A shows the
data/model ratio found when a simple ``diskbb'' plus power-law model
is used to fit the data, panel B shows the ratio to a ``diskpn'' plus
power-law model, panel C shows the ratio to a ``diskbb'' plus
``CompTT'' model, and panel D shows the ratio to the ``bulk-motion
Comptonization'' model (see the text, and Miller et al.\ 2006a). }
\label{label1}
\end{figure}

Finally, it is sometimes argued that -- whatever the true spectral
continuum -- partial covering of the central engine might act to
falsely create the appearance of a broad iron line by adding opacity
in ionized K-shell absorption edges.  The high signal-to-noise ratio
achieved in long observations of stellar-mass black holes makes it
possible to test this possibility.  Figure 4 shows the {\it
XMM-Newton} spectrum of GX~339$-$4 described in Miller et al.\
(2004b), fit with a disk plus power-law model, and allowing two edges
to float in the Fe~K edge band (7.1--9.3~keV) instead of adding a
relativistic iron line and disk reflection.  The spectrum is not well
fitted by a model consistent with partial covering; indeed, the model
is also significantly worse in a statistical sense.  Partial covering
also appears to be a significantly worse description of Seyfert-1
spectra where the high energy continuum is well-defined (Reynolds et
al.\ 2004).

\begin{figure}
\includegraphics[width=55mm,height=83mm,angle=-90]{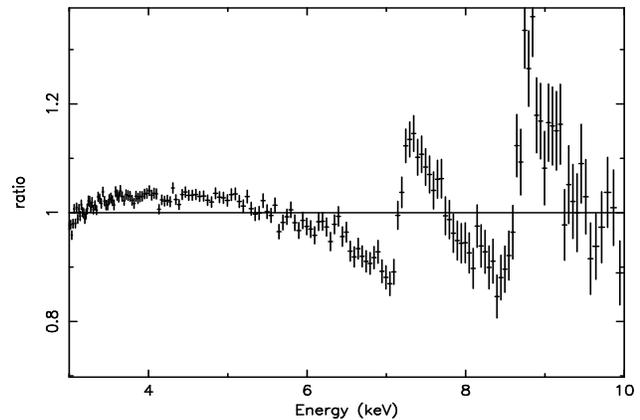}
\caption{The data/model ratio shown above is the result of fitting an
absorbed ``diskbb'' plus power-law continuum to the {\it
XMM-Newton}/EPIC-pn spectrum of GX~339$-$4 in a bright state.  Two
edges were allowed to float between 7.1--9.3~keV, as per a scenario in
which patchy absorption partially covers the central engine.  The
resulting fit shown above is clearly unacceptable, and gives a
significantly worse fit statistic than fits with a relativistic line
model ($\chi^{2}/\nu = 5913.5/1899$, versus $\chi^{2}/\nu =
3456.5/1894$, Miller et al.\ 2004b).  In good spectra from stellar-mass
black holes, partial covering can be ruled-out, further demonstrating
that relativistic line profiles are robust.}
\label{label1}
\end{figure}

\section{Methods of Constraining Spin}

The frequency of high-frequency QPOs may be related to black hole
spin, and most models for high-frequency QPOs point to high spin
parameters, especially when the QPOs are seen in a 2:3 ratio.  At
present, however, a modulation mechanism has not been uniquely
identified.  In contrast, the mechanism by which relativistic iron
lines are produced and shaped is well understood.

It is also possible to constrain the spin of a stellar-mass black hole
by measuring the X-ray continuum flux from the accretion disk.  In
practice, this is complicated by many factors, including: the
accuracy with which the ISM absorption is known, the assumed form of
the hard component, the accuracy with which the distance and
inclination of the system and mass of the black hole are known, the
assumed inner disk torque prescription, the mass accretion rate
through the disk (which contains an uncertainty in the $\alpha$
viscosity parameter), and the accuracy with which spectral hardening
through the disk atmosphere can be characterized.

As an example of the way in which estimates of black hole spin based
on the disk continuum are subject to considerable systematic
uncertainty, we consider an {\it RXTE}/PCA spectrum of the 4U
1543$-$475 (observation 70133-01-06-00, which started on 2002-06-29 at
14:08:16 TT).  Shafee et al.\ (2006) previously fit the spectrum from
this observation with a new disk model, having identified it as being
a true soft state spectrum.  4U 1543--475 is a better choice for this
test than, e.g., GRS~1915$+$105 (found to have a high spin parameter
by McClintock et al.\ 2006 based on disk fits) as 4U 1543$-$475 has a
much lower column density, which considerably reduces one source of
systematic difficulty in measuring the disk continuum.

We used the standard PCU spectrum available from the {\it RXTE} archive
in the 2.8--25.0~keV band.  To further reduce a systematic
uncertainty, the column density was fixed at $N_H = 4.0\times
10^{21}~{\rm cm}^{-2}$ as per Dickey \& Lockman (1990).  We first fit
this spectrum with a simple ``diskbb'' plus power-law model, and next
with a ``diskbb'' plus ``CompTT'' model (with the initial temperature
set to that of the disk, and the coronal temperature fixed to 50~keV).
These models give statistically equivalent fits, and both models yield
a disk color temperature of $kT = 0.89$~keV.  However, they imply very
different disk fluxes, and it is the flux that determines the emitting
area and thereby the proximity of the disk to the black hole. Figure 5
shows that even when fitting spectra obtained in a soft disk-dominated
state, different models for the hard component can generate a factor
of 2 difference in the disk flux inferred at 3~keV.

\begin{figure}
\includegraphics[width=55mm,height=83mm,angle=-90]{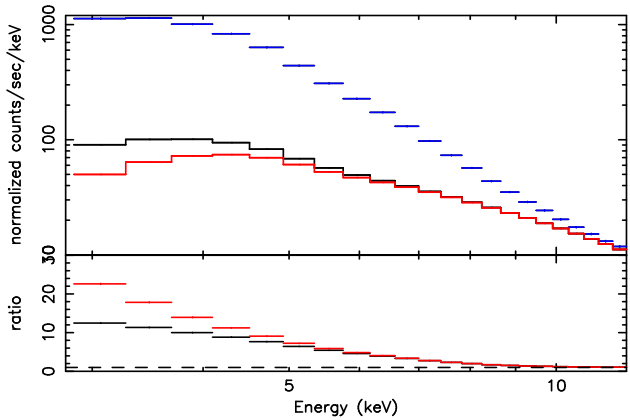}
\caption{Two statistically equivalent fits to a spectrum of 4U
1543$-$475 (shown in blue) are shown above (the disk flux was later set to
zero to make this figure).  A model consisting of ``diskbb'' and power-law
components is shown in black, and a model consisting of ``diskbb'' and
``CompTT'' components is shown in red.  Even when the hard flux is
minimal, the choice of hard component can strongly affect the disk
flux that is measured, and efforts to constrain the spin of the black
hole via the disk continuum.  The data/model ratio above shows that
the disk flux at 3~keV differs by a factor of $\sim$2, due only to the
different models used to fit the hard X-ray flux.}
\label{label1}
\end{figure}

We next fit the spectrum of 4U~1543$-$475 with the new variable-spin
disk model considered in Shafee et al.\ (2006) and McClintock et al.\
(2006), using the same power-law model for the hard component.  The
``Kerrbb'' model (Li et al.\ 2005) includes the spin of the black hole
as a variable parameter, but requires one to know a torque/accretion
power dissipation ratio, the inclination of the inner disk, the mass
of the black hole, the mass accretion rate (which depends on the
torque to accretion dissipation ratio), the distance to the black
hole, the spectral hardening factor for transfer through the disk
atmosphere, whether or not disk self-irradiation is important (a
binary switch), and whether or not limb-darkening should be included
(also a binary switch).  In our fits, we assumed the default switches
(a torque/accretion dissipation ratio of zero, inclusion of
self-irradiation, and zero limb darkening), the mass of the black hole
was constrained to $9.4\pm 2.0 M_{\odot}$, its distance was
constrained to $7.5\pm 1.0$~kpc, and the inclination was fixed to 20.7
degrees, as per Park et al.\ (2004).  We find that fits with $a=0.9$
and $a=0.3$ are statistically equivalent for a small range in mass
accretion rate.  Clearly, constraints on the black hole spin parameter
using the disk continuum are strongly affected by the choice of the
hard component, and by the assumed parameters in the disk model.

\begin{figure}
\includegraphics[width=55mm,height=83mm,angle=-90]{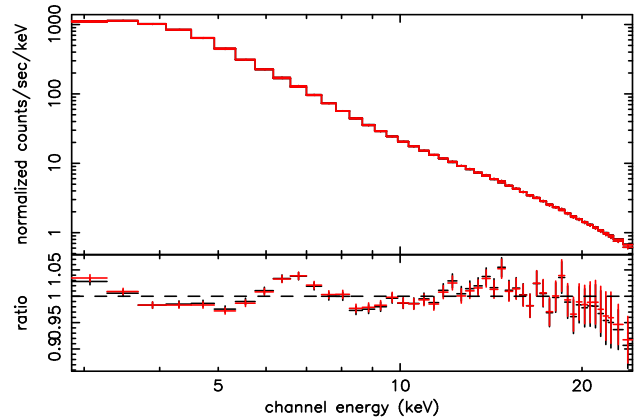}
\caption{Two examples of the same spectrum of 4U 1543$-$475 are shown
above.  Both examples are fit with a model including the ``Kerrbb''
disk and a power-law components.  The model shown in black was fit
with $a=0.9$ fixed, and the model in red was fit with $a=0.3$ fixed
(please see the text).  The data/model ratio clearly shows that both
continuum fits are acceptable, and statistically equivalent.  Indeed,
the difficulties inherent in obtaining robust black hole spin
constraints using the disk continuum as illustrated here motivated
efforts to use Fe~K emission lines and QPOs for this purpose.}
\label{label1}
\end{figure}

In summary, relativistic iron lines are not the only way of
constraining spin in stellar-mass black holes, but they present a few
specific advantages over other methods.  The examination in this and
the previous section shows that lines are relatively free of
systematic difficulties, at least when compared to QPO and disk
continuum spin constraints.  The largest source of uncertainty for
relativistic iron lines is the extent to which Comptonization might
broaden the line.  It has been shown that for Comptonization to
contribute significantly, however, the scattering region must be as
optically thick as the disk itself (e.g., Reynolds \& Wilms 2000),
which is unphysical.  The inclination of the inner accretion disk is
important for both relativistic iron lines and disk continua, but spin
constraints based on iron lines do not require one to know the mass of
a given black hole or its distance, nor do such constraints rely upon
the many other parameters important for accurate disk continuum fits.
Both disk line and disk continuum diagnostics assume that emission
from within the innermost stable circular orbit is negligible, which
is reasonable if the material in the plunging region is hot and
optically-thin.

\section{Implications for AGN and Future Directions}

The high signal-to-noise across the broad X-ray bandpass afforded by
the high flux observed from stellar-mass black holes permits robust
fits to relativistic Fe~K lines in medium and high-resolution X-ray
spectra.  Relativistic line studies can explore General Relativistic
effects near to black holes, the spin evolution of stellar-mass black
holes, and can bear on the role of black hole spin in launching
relativistic jets.  Beyond these considerations, stellar-mass black
hole spectra are an excellent point of comparison for X-ray spectra
obtained from Seyfert-1 AGN.  Whereas the nature of the low and high
energy continuum in Seyfert-1 X-ray spectra is sometimes ambiguous,
and whereas partial covering of the central engine (which might cause
the false appearance of a relativistic line) can be hard to exclude,
in stellar--mass black holes it is possible to show that strong
relativistic lines are robust against such effects (see above).  The
powerful combination of {\it XMM-Newton} and {\it RXTE} spectra has
been instrumental in demonstrating the robustness of relativistic
lines in stellar-mass black holes.

As noted above, the best relativistic spectra obtained from
stellar-mass black holes and Seyfert-1 AGN can be fit using the same
broad-band disk reflection model (Miller et al.\ 2004b).  Such
findings serve to indicate that the ionization of the inner accretion
disk is the only major difference between the inner accretion flows.
Given these results, and the robust and widespread nature of
relativistic lines in stellar-mass black holes, it is reasonable to
expect that relativistic lines should be common in Seyfert-1 AGN.
Low-energy absorption and poor sensitivity in the energy range above
the Fe~K band may hinder strong detections of relativistic lines in
soft X-ray spectrometers, but if stellar-mass black holes can drive
our expectations then such lines should be fairly common.  There are
already indications that improved broad-band sensitivity alone
provides stronger detections of relativistic lines in Seyfert-1 AGN
(see the contributions by Nandra, Reeves, and Markowitz in this
proceedings).

The near-term future for relativistic line studies is very promising,
and a few directions may be especially important to obtaining improved
spin constraints, evidence of related relativistic phenomena, and the
nature of orbits close to black holes.

At present, only two relativistic line models are commonly fit: the
``diskline'' model (Fabian et al.\ 1989) and the ``Laor'' model (Laor
1991).  The former describes lines originating around a zero-spin
Schwarzschild black hole, and the latter describes line emission
originating around a maximal-spin Kerr black hole.  The vast majority
of broad iron lines are better described by the Kerr model; this fact,
and the simple FWHM obtained when line profiles are fit with a
Gaussian function, are already a strong indications that most black
holes may have high spin parameters.  If one makes the approximation
that the line profiles expected at high spin parameters do not differ
very greatly, then one can associate the inner radius measured using
the Laor model with the innermost stable circular orbit, and derive a
more conservative estimate of the black hole spin parameter.
Recently, models for relativistic lines with spin as a variable
parameter have been calculated, and are available for spectral fitting
in standard packages.  These models are called ``kdline'' (Beckwith \&
Done 2004, 2005) and ``ky'' (Dovciak, Karas, \& Yaqoob 2004).  A third
such model (not yet public) was used to fit the line profile
in MCG--6-30-15, and found a spin that is broadly consistent with some
estimates using the Laor model ($a \geq 0.987$, Brenneman \& Reynolds
2006).  In the near future, it is vital that similar fits be made to a
number of broad Fe K line profiles, both in Seyfert-1 AGN and
stellar-mass black holes.

In AGN, robust QPOs have not yet been detected, and measuring the disk
continuum is even more difficult than in stellar-mass black holes.
The possibility of constraining spin via independent diagnostics, then,
is unique to stellar--mass black holes.  Though each method of
constraining spin has advantages and disadvantages, the
best way forward for stellar-mass black holes will be to bring each
diagnostic to bear simultaneously, and in as many sources as possible.  
 
Studies of Fe~K line and continuum variability are another important
direction for the future.  As noted above, a model including the
effects of relativistic light bending close to a spinning black hole
predicts a specific pattern of line and continuum variability
(Miniutti et al.\ 2003).  The variability patterns seen in some
Seyfert-1 AGN and one stellar mass black hole appear to follow this
pattern (Miniutti, Fabian, \& Miller 2004; Rossi et al.\ 2005).  Until
recently, most studies of Fe~K lines in both Seyfert-1 AGN and
stellar-mass black hole have focused on establishing the presence of
relativistic lines in time-averaged spectra, and in measuring the
parameters of the time-averaged line profile.  Long observations
already made with {\it ASCA}, {\it BeppoSAX}, {\it Chandra}, and {\it
XMM-Newton} offer the possibility of further testing this model with
Seyfert-1 AGN.  In the case of stellar-mass black holes, long
observations with these observatories, but also with {\it RXTE}, offer
the same opportunity.  Indeed, {\it RXTE} has obtained hundreds of
observations of many stellar-mass black holes in outburst, but the
line versus continuum flux has only been examined in one case so-far.
Light bending should occur near to black holes, especially those with
high spin parameters, and variability studies may be able to reveal
this effect.

Finally, searches for periodic and quasi-periodic signatures in
relativistic iron line profiles and disk reflection hold extraordinary
promise.  The orbital timescales in the inner disk around supermassive
black holes can now be probed with {\it XMM-Newton} and {\it Suzaku}.
Nearly periodic line variability consistent with a flare orbiting the
black hole at 7--16~$GM/c^{2}$ was detected in a long observation of
NGC 3516 (Iwasawa, Miniutti, \& Fabian 2004).  A similar periodicity
has also been reported in long {\it XMM-Newton} observation of Mrk 766
(Turner et al.\ 2006).  The statistical significance of these periods
is modest, but future long observations can prove decisive.  In
stellar-mass black holes, the equivalent width and flux of Fe~K
emission lines in the spectrum of GRS 1915+105 has recently been found
to vary with QPO phase (Miller \& Homan 2005).  This can be explained
in terms of precession in the inner disk (Schnittman, Homan, \& Miller
2006), and may hint at Lense-Thirring precession.  While it is seldom
that a QPO is observed with the high rms amplitude required for such
studies, this is again an area in which future observations with {\it
RXTE}, {\it XMM-Newton}, and {\it Suzaku} can reveal more information
on orbits close to black holes.

In summary, comparing and contrasting relativistic Fe~K lines and disk
reflection in Seyfert-1 AGN and stellar-mass black holes is extremely
valuable, and continues to yield important insights.  There is a
bright future for studies of lines in stellar-mass black holes using
the present array of X-ray observatories; significant progress is
within reach.  In the long run, the insights that relativistic lines
in stellar-mass black hole spectra provide strongly argue in favor of
retaining a bright (1-2 Crab) source capacity in future X-ray missions
such as {\it Constellation-X} and {\it XEUS}.

\acknowledgements
I wish to acknowledge Chris Reynolds and Andy Fabian for comments on
this contribution.  I would like to thank Norbert Schartel, Andy
Fabian, Arvind Parmar, Matthias Ehle, Maria Diaz-Trigo, and the
XMM-Newton staff responsible for arranging an excellent conference.

\newpage%%%%%%%%%%%%%%%%%%%%%%%%%%%%%%%%%%%%%%%%%%%%%%%%%%%%%%

\end{document}